\documentclass[final,10pt,journal,letterpaper]{IEEEtran}

\newenvironment{Ni}[1]{\textcolor{red}{#1}}

\usepackage[utf8]{inputenc}
\usepackage[T1]{fontenc}
\usepackage{url}
\usepackage{ifthen}
\usepackage{cite}
\usepackage[cmex10]{amsmath}
\usepackage{amssymb,amsthm}
\usepackage{dsfont}
\usepackage{mathtools}
\usepackage{amsmath}
\usepackage{xfrac}
\usepackage{stmaryrd}
\usepackage{tikz}
\usetikzlibrary{automata, arrows}

\DeclareMathOperator{\argmin}{arg\,min}

\newtheorem{theorem}{Theorem}

\newtheorem{definition}{Definition}
\newtheorem{proposition}[theorem]{Proposition}

\newtheorem{remark}{Remark}

\newtheorem{corollary}[theorem]{Corollary}

\newcommand{\range}[1]{\llbracket #1 \rrbracket}

\begin{document}
\title{Measuring Information Leakage in \\ Non-stochastic Brute-Force Guessing}

\author{Farhad Farokhi,~\IEEEmembership{Senior Member,~IEEE, }Ni Ding,~\IEEEmembership{Member,~IEEE }\thanks{F. Farokhi and N. Ding are with the University of Melbourne. When working on this paper, F. Farokhi was also affiliated with CSIRO's Data61.} \thanks{emails:\{farhad.farokhi,ni.ding\}@unimelb.edu.au}\thanks{The work of F. Farokhi is, in part, funded by the Melbourne School of Engineering at the University of Melbourne.}
\thanks{The work of Ni Ding is funded by the Doreen Thomas Postdoctoral Fellowship at the University of Melbourne.}}

\maketitle

\begin{abstract} We propose an operational measure of information leakage in a non-stochastic setting to formalize privacy against a brute-force guessing adversary. We use uncertain variables, non-probabilistic counterparts of random variables, to construct a guessing framework in which an adversary is interested in determining private information based on uncertain reports. We consider brute-force trial-and-error guessing in which an adversary can potentially check all the possibilities of the private information that are compatible with the available outputs to find the actual private realization. The ratio of the worst-case number of guesses for the adversary in the presence of the output and in the absence of it captures the reduction in the adversary's guessing complexity and is thus used as a measure of private information leakage. We investigate the relationship between the newly-developed measure of information leakage with the existing non-stochastic maximin information and stochastic maximal leakage that are shown arise in one-shot guessing.
\end{abstract}

\section{Introduction}



Recently, maximal leakage based on {one-shot} guessing~\cite{issa2018operational} and guessing leakage based on brute-force guessing~\cite{osia2019privacy} have been developed to provide operational information-leakage metrics for privacy analysis. These notions have started a new wave of research in information-theoretic privacy with interpretable or operational measure of private information leakage~\cite{liao2018tunable,li2018privacy}. {In some cases, however, probability distributions of the underlying variables or conditional probability of outputs given private data might not be known \textit{a priori} or might change unpredictably over time. For instance, when considering small datasets, enough data might not be available to make probabilistic inference about the population and, thus, we may want to investigate whether an adversary can gain private information that is not based on statistics. Alternatively, we may need to avoid randomized policies for privacy preservation.} For instance, this could be due to concerns about un-truthfulness in privacy-preserving reports~\cite{bild2018safepub,Poulis2015} or complications in financial auditing and fraud detection~\cite{bhaskar2011noiseless, nabar2006towards}. {Therefore}, {in these cases}, there is a need to investigate information leakage in non-stochastic frameworks.

In this paper, we propose a measure of information leakage in a non-stochastic framework. We do so to provide an interpretation for the recent results on non-stochastic privacy~\cite{8662687,ding2019developing,farokhi2019noiseless}. We use uncertain variables, non-stochastic counterparts of random variables introduced in~\cite{nair2013nonstochastic}, to construct a guessing framework in which an adversary is interested in determining private information based on available outputs. We consider a brute-force guessing setup in which an adversary potentially checks all the possibilities of the private information that are compatible with the outputs to find the actual realization of the private information. This is similar to the interpretation of~\cite{issa2018operational} for password guessing or side-channel attacks on cipher systems in which an adversary can repeatedly checks all the possible combinations that are compatible with its observations. However, the approach of~\cite{issa2018operational} is based on the probability of successful deduction/inference with just one guess while we use the number of guesses in a repeated scenario. This is similar to the brute-force guessing framework in~\cite{osia2019privacy} with the exception of {avoiding} distributions or statistics. The ratio of the worst-case number of guesses for the adversary in the presence of the outputs and in the absence of them captures the reduction in the adversary's guessing complexity and is thus used as a measure of information leakage. 

{Although a brute-force interpretation of leakage is used in this paper, we follow the axioms}\footnote{Not all the requirements in~\cite{issa2018operational} are axioms, e.g., the requirement for the leakage to accord with intuition, but most can be regarded as fundamental properties required for private information leakage metric.} of~\cite{issa2018operational} for guiding the development of the information leakage metric. These axioms are, in fact, relevant to any notion of information leakage. Therefore, we require that the introduced information-leakage metric (R1) explain leakage in an operational manner (what bounding leakage means in practice), (R2) require minimal assumptions about the privacy-intrusive adversary, (R3) satisfy properties, such as (R3.a) data-processing inequality (post processing does not increase leakage), (R3.b) independence property (independent outputs result in zero leakage), and (R3.c) additivity property (akin to composition rule in differential privacy), and finally, (R4) accord with intuition. 

In summary, this paper makes the following contributions:
\begin{itemize}
    \item Proposing a non-stochastic brute-force guessing framework for measuring information leakage in which the ratio of the worst-case number of guesses for the adversary in the presence of the output and in the absence of it is used to capture the reduction in the adversary's guessing  complexity and to define a measure of information leakage;
    \item Measuring leakage from the private data to the outputs when we are aware of adversary's intentions (i.e., what {sensitive attribute/data} it wants to guess) and when we are not aware of the adversary's intentions, {which} is defined based on the maximal information leakage; 
    \item Demonstrating that the non-stochastic brute-force leakage satisfies the axioms outlined for information leakage in~\cite{issa2018operational}, such as operational interpretation, minimality of assumptions on the adversary, data-processing inequality, independence property, and additivity;
    \item Presenting identifiability{,} a new notion of privacy based on the developed maximal measure of information leakage{,} in this paper;
    \item Relating the non-stochastic information leakage based on the presented brute-force guessing framework to maximin information~\cite{nair2013nonstochastic}, which we prove that stems naturally from one-shot guessing with perfect accuracy, and stochastic maximal leakage, which is shown to relate to stochastic one-shot guessing~\cite{issa2018operational}.
\end{itemize}


The rest of the paper is organized as follows. We present preliminary material on uncertain variables and non-stochastic information theory in Section~\ref{sec:uv}. In Section~\ref{sec:info_leakage}, we present the measure of information leakage {of a specific sensitive attribute to the output and use it as the privacy measure when we are aware of adversary's intentions}. In Section~\ref{sec:maximal_leakage}, we extend this notion to when we are not aware of the adversary's intentions by defining maximal non-stochastic brute-force leakage. We present non-stochastic identifiability as a new notion of non-stochastic privacy in Section~\ref{sec:privacy}. We compare the brute-force notion of non-stochastic information leakage with one-shot guessing measures, such as maximin information and stochastic maximal leakage in Section~\ref{sec:relationship}. Finally, we conclude the paper in Section~\ref{sec:conc}.

\section{Uncertain Variables} \label{sec:uv}
We borrow the following concepts from~\cite{nair2013nonstochastic}. Consider uncertainty set $\Omega$. An uncertain variable, uv in short, is a mapping on $\Omega$. For example, for uv $X:\Omega\rightarrow\mathbb{X}$, $X(\omega)$ is the realization of uv $X$ corresponding to uncertainty $\omega\in \Omega$. For any two uvs $X$
and $Y$, the set $\range{X,Y}:=\{(X(\omega), Y(\omega)):
\omega\in\Omega \}\subseteq \range{X}\times \range{Y}$ is their \textit{joint range}. For uv $X$, $\range{X}:=\{X(\omega):\omega\in\Omega \}$ denotes its \textit{marginal range}.
The \textit{conditional range} of uv $X$, conditioned on realizations of uv $Y$ belonging to the set $\mathcal{Y}$, is $\range{X|Y(\omega)\in \mathcal{Y}}:=\{X(\omega):\exists \omega\in\Omega  \mbox{ such that } Y(\omega)\in\mathcal{Y}\} \subseteq \range{X}.$ If $\mathcal{Y}=\{y\}$ is a singleton,  $\range{X|Y(\omega)\in \{y\}}=\range{X|Y(\omega)\in \mathcal{Y}}$ is replaced with $\range{X|Y(\omega)=y}$ or $\range{X|y}$ when it is clear from the context. For any two uvs $X$ and $Y$, we define the notation  $\range{Y|X}:=\{\range{Y|X(\omega)=x},\forall x\in\range{X}\}$. We sometimes refer to $\range{Y|X}$ as a {non-stochastic} channel as $\range{Y|X}$ fully {characterizes the non-stochastic} communication channel from $X$ to $Y$. In this paper, we only deal with \textit{discrete} uvs possessing finite\footnote{Extension to countably infinite sets is straightforward with extra care when manipulating extended real numbers (i.e., infinity).} ranges.

Uvs $X_1$ and $X_2$ are unrelated if $\range{X_1|X_2(\omega)=x_2}=\range{X_1}$ for all $x_2\in\range{X_2}$ and \textit{vice versa}.  Similarly, $X_1$ and $X_2$ are conditionally unrelated given $Y$ if $\range{X_1|X_2(\omega)=x_2,Y(\omega)=y}=\range{X_1|Y(\omega)=y}$ for all $ (x_2,y)\in\range{X_2,Y}$. Uvs $X_i$, $i=1,\dots,n$, are \emph{unrelated} if $\range{X_1,\dots,X_n}= \range{X_1}\times \cdots \times \range{X_n}$ and \emph{conditionally unrelated} given $Y$ if $\range{X_1,\dots,X_n|Y(\omega)=y} =\range{X_1|Y(\omega)=y}\times \cdots \times \range{X_n|Y(\omega)=y}$ for all $y\in\range{Y}$. Uvs $X$, $Y$, and $Z$ form a Markov (uncertainty) chain, denoted by $X-Y-{Z}$, if $X$ and $Z$ are unrelated conditioned on $Y$, that is, $\range{X|Z(\omega)=z,Y(\omega)=y} =\range{X|Y(\omega)=y}$ for all $(z,y)\in\range{Z,Y}.$ Note that, by symmetry of the definition of unrelated uvs, $X-Y-Z$ forms a Markov chain if and only if $Z-Y-X$ forms a Markov chain. We say $X_1-X_2-\cdots-X_n$ forms a Markov chain if $X_i-X_j-X_\ell$ forms a Markov chain for any $1\leq i {<} j {<} \ell\leq n$.

Non-stochastic entropy of uncertain variable $X$ is defined as $H_0(X):=\log_2(|\range{X}|).$ This is often described as the Hartley entropy~\cite{hartley1928transmission, nair2013nonstochastic}, which coincides with the R\'{e}nyi entropy of order $0$ for discrete variables~\cite{sason2017arimoto,renyi1961measures}.
%
Conditional (or relative) entropy of uv $X$ given $Y$ is given by
$H_0(X|Y):=\max_{y\in\range{Y}}\log_2(|\range{X|Y(\omega)=y}|)$. This is the Arimoto-R\'{e}nyi conditional entropy of order $0$~\cite{sason2017arimoto,10022581674}. Based on this, we can define $I_0(X;Y):=H_0(X)-H_0(X|Y)$. This is equivalent to the $0$-mutual information~\cite{sason2017arimoto,verdu_alpha_mutual_2015}.
%

We end this section by presenting the definition of maximin information from non-stochastic information theory~\cite{nair2013nonstochastic}. Consider uvs $X$ and $Y$. Any $x,x'\in\range{X}$ are $\range{X|Y}$-overlap connected if there exists a finite sequence of conditional ranges $\{\range{X|Y(\omega)=y_i}\}_{i=1}^n$ such that $x\in\range{X|Y(\omega)=y_1}$, $x'\in\range{X|Y(\omega)=y_n}$, and $\range{X|Y(\omega)=y_i}\cap \range{X|Y(\omega)=y_{i+1}}\neq \emptyset$ for all $i=1,\dots,n-1$. We say $\mathcal{A}\subseteq\range{X}$ is $\range{X|Y}$-overlap connected if all $x,x'\in\mathcal{A}$ are $\range{X|Y}$-overlap connected. Further, $\mathcal{A},\mathcal{B}\subseteq\range{X}$ are $\range{X|Y}$-overlap isolated if there does not exist $x\in\mathcal{A},x'\in\mathcal{B}$ such that $x,x'$ are $\range{X|Y}$-overlap connected.  An $\range{X|Y}$-overlap partition is a partition of $\range{X}$ such that  each member set is $\range{X|Y}$-overlap connected and any two member sets are $\range{X|Y}$-overlap isolated. There always exists a unique $\range{X|Y}$-overlap partition~\cite{nair2013nonstochastic}, which is denoted by $\range{X|Y}_\star$. The maximin information is $I_\star(X;Y):=\log_2(|\range{X|Y}_\star|).$
In~\cite{nair2013nonstochastic}, it is proved that  $|\range{X|Y}_\star|=|\range{Y|X}_\star|$ and thus $I_\star(X;Y)=I_\star(Y;X)$. The overlap partition captures common uv~\cite{mahajanrelationship}, an extension of common random variable~\cite{wolf2004zero} to uvs. This relationship explains the relationship between entropy of the common uv, which is equal to the maximin information, and the zero-error capacity~\cite{wolf2004zero,nair2013nonstochastic}.



\section{Information Leakage in Brute-Force Guessing}  \label{sec:info_leakage}

Consider uv $X$ containing sensitive data $U$, which is interpreted as some attribute or feature of $X$ that is computable by some function $g:\range{X}\rightarrow\range{U}$, i.e., $U=g\circ X$. Note that, by construction, $|\range{U}|\leq |\range{X}|$. Let $Y$ be an observable uv that depends on $X$, e.g., $X$ and $Y$ are the input and output, respectively, of a (privacy-preserving) channel.\footnote{The conditional range $\range{Y|X}$ characterizes this channel, which can also be regarded as a non-stochastic privacy-preserving scheme.} These uvs forms a Markov chain $U-X-Y$. An adversary wants to guess $U$ correctly given $Y$. For instance, consider an example in which $X$ captures weight and height of an individual, and $U$ denotes body mass index. In such an example, insurance agencies might be interested in deducing the body mass index of an individual (due to its correlation with heart disease) based on publicly released data $Y$ while they do not have any particular interest in learning an individual's height and weight separately.

We assume that the adversary can guess the value of $U$ in a brute-force trial-and-error manner. That is, the adversary chooses a distinct element $u\in\range{U}$ each time and tests\footnote{We assume that the adversary has access to an oracle that can determine whether $U(\omega)$ is equal to $u$ (for a given $u\in\range{U}$) or not.} whether the actual value $U(\omega)$ equals $u$. The adversary repeats this procedure until the answer is `yes'. We consider the number of trials before the successful guess. Without observations of $Y$, the adversary must try at most $|\range{U}|$ times. However, with access to observation $Y(\omega)=y \in \range{Y}$, the actual value of $U(\omega)$ lies in the conditional range $\range{U|Y(\omega)=y}$ and therefore the maximum number of trials is $|\range{U|Y(\omega)=y}|$. {Since} the number of trials {is} proportional to the inference cost/effort of the adversary, the ratio $|\range{U}|/|\range{U|Y(\omega)=y}|$ captures the reduction in the adversary's maximum cost for guessing $U$ upon the observation $\range{U|Y(\omega)=y}$. This coincide{s} with the definition of the information gain $\log_2(|\range{U}|/|\range{U|Y(\omega)=y}|)$ in~\cite{kolmogorov1959varepsilon}, where $\log_2(|\range{U|Y(\omega)=y}|)$ denotes the `combinatorial' conditional entropy. The adversary's reduction in guessing cost can be interpreted as the information gained about uv $U$ from {the} observation $Y(\omega)=y$.

Note that the measure $\log_2(|\range{U}|/|\range{U|Y(\omega)=y}|)$ is also consistent with the \emph{stochastic brute-force guessing leakage} $H_G(U) - \mathbb{E}_{Y}[H_G(U|Y(\omega)=y)]$
proposed in~\cite[Definition~3]{osia2019privacy} for rvs $U$ and $X$. This measure is based on the guessing entropy\footnote{The guessing entropy $H_G(U)$ denotes the minimum average number of trials for guessing the realization of $U$. This results from the optimal brute-force guessing strategy of the adversary to pick $u_i\in \range{U}$, i.e., the element in $\range{U}$ with the $i$-th largest probability $\mathbb{P}\{U(\omega)=u_i\}$, at the $i$-th trial~\cite{massey1994guessingentropy}. } in~\cite{massey1994guessingentropy} defined as $H_G(U):= \sum_{i=1}^{|\range{U}|} i \mathbb{P}\{U(\omega)=u_i\}$, where $(u_i)_{i=1}^{|\range{U}|}$ are such that $\mathbb{P}\{U(\omega)=u_1\} \geq \mathbb{P}\{U(\omega)=u_2\} \geq \dotsc \geq \mathbb{P}\{U(\omega)=u_{|\range{U}|}\}$.
Similarly, the conditional guessing entropy is $H_{G}(U | Y(\omega)=y) = \sum_{i=1}^{|\range{U|Y(\omega)=y}|} i \mathbb{P}\{U(\omega)=\tilde{u}_i|Y(\omega)=y\}$ for each $y \in \range{Y}$, where $(\tilde{u}_i)_{i=1}^{|\range{U|Y(\omega)=y}|}$ are such that $\mathbb{P}\{U(\omega)=\tilde{u}_1|Y(\omega)=y\} \geq \mathbb{P}\{U(\omega)=\tilde{u}_2|Y(\omega)=y\} \geq \dotsc \geq \mathbb{P}\{U(\omega)=\tilde{u}_{|\range{U|Y(\omega)=y}|}|Y(\omega)=y\}$. When there is no $\sigma$-field or probability measure over $\range{U}$, $H_{G}(U)$ and $H_G(U|Y(\omega)=y)$ reduce to the prior guessing cost $\log_2(|\range{U}|)$ and posterior guessing cost $\log_2(|\range{U|Y(\omega)=y}|)$, respectively, {by} replacing the {expectation} with the worst-case. To quantify the \emph{non-stochastic brute-force guessing leakage}, we consider the difference between $\log_2(|\range{U}|)$ and the minimum guessing cost $\min_{y \in \range{Y}}\log_2(|\range{U|Y(\omega)=y}|)$ as follows.

\begin{definition}[\underline{Non-Stochastic Brute-force Guessing Leakage}]
For a given uv $U$, the non-stochastic leakage from $U$ to $Y$ is
\begin{align*}
\mathcal{L}(U\rightarrow Y)=&\log_2
\left(\frac{|\range{U}|}{\displaystyle\min_{y\in \range{Y}} |\range{U|Y(\omega)=y}|} \right)
\\
=&
 \max_{y\in \range{Y}} \log_2
\left(\frac{|\range{U}|}{|\range{U|Y(\omega)=y}|} \right).
\end{align*}
\end{definition}

The measure $\mathcal{L}(U\rightarrow Y)$ quantifies the maximum reduction in the guessing cost of the adversary after observing $Y$, which indicates the most information gained by the adversary in the sense of~\cite{kolmogorov1959varepsilon}. This measure has been previously used as the non-stochastic information leakage in~\cite{8662687,ding2019developing} for privacy analysis, e.g., in the case of $k$-anonymity~\cite{8662687}. Hence, this definition provides an operative meaning to the non-stochastic information leakage and can be used as its interpretation for privacy analysis.

In the following proposition, we show that non-stochastic leakage satisfies the data-processing inequality. This implies that, for a given uv $X$ and a specified attribute $U$ of $X$, the leakage is non-increasing along cascading channels $\range{Y|X}$ and $\range{Z|Y}$. This is in line with axiom R3.a of an operational notion of information leakage in \cite{issa2018operational}. This is an important requirement as it shows that a curator does not need to worry about an increased risk {incurred by any post processing} after releasing outputs.



\begin{proposition}[\underline{Data Processing Inequality}] If Markov chain $U-X-Y-Z$ holds, $\mathcal{L}(U\rightarrow Z)\leq \mathcal{L}(U\rightarrow Y)$.
	\label{prop:DPI_1}
\end{proposition}

\begin{IEEEproof} Note that
	\begin{align*}
	\range{U|Z(\omega)=z}
	&=\bigcup_{y\in\range{Y}} \range{U|Z(\omega)=z,Y(\omega)=y}\\
	&=\bigcup_{y\in\range{Y}:(y,z)\in\range{Y,Z}} \range{U|Y(\omega)=y},
	\end{align*}
	where the last equality follows from that $U-Y-Z$ is  a Markov chain, i.e., $U$ and $Z$ are unrelated given $Y$.
	Notice that $y\in\range{Y}$ and $(y,z)\in\range{Y,Z}$ implies that $y\in\range{Y|Z(\omega)=z}$. As a result,
	\begin{align} \label{eqn:proof:1}
	\range{U|Z(\omega)=z}
	&=\bigcup_{y\in\range{Y|Z(\omega)=z}} \range{U|Y(\omega)=y}.
	\end{align}
	Let $z^* \in \argmin_{z\in\range{Z}} |\range{U|Z(\omega)=z}|$. For any $y^*\in\range{Y|Z(\omega)=z^*}$, $\range{U|Y(\omega)=y^*} \subseteq \range{U|Z(\omega)=z^*} $ because of~\eqref{eqn:proof:1}. Hence,
	\begin{align*}
	\min_{y\in\range{Y}} |\range{U|Y(\omega)=y}|
	&\leq |\range{U|Y(\omega)=y^*}|\\
	&\leq |\range{U|Z(\omega)=z^*}|\\
	&=\min_{z\in\range{Z}}|\range{U|Z(\omega)=z}|,
	\end{align*}
	which, because of the monotonicity of the logarithm, gives rise to the inequality $\mathcal{L}(U\rightarrow Z)\leq \mathcal{L}(U\rightarrow Y)$.
\end{IEEEproof}

The following result shows that the non-stochastic brute-force guessing leakage is a measure of relatedness between two uvs. In fact, the leakage is equal to zero if two uvs are unrelated. Evidently, the most private case arises from ensuring that $X$ and $Y$ are unrelated. In this case, the realizations of $Y$ do not provide any useful information about $X$ or its derivatives, e.g., $U$.  This is again in line with axiom R3.b of an operational notion of information leakage~\cite{issa2018operational}.


\begin{proposition}[\underline{Bounding Leakage}]
	\label{prop:Bounding_Leakage}
	 $ \mathcal{L}(U\rightarrow Y)\geq 0$ with equality if $X$ and $Y$ are unrelated.
\end{proposition}
\begin{IEEEproof} The inequality follows from that $\range{U|Y(\omega)=y}\subseteq \range{U}$ and, as a a result, $|\range{U}|/|\range{U|Y(\omega)=y}|\geq 1$. If $X$ and $Y$ are unrelated, $U$ and $Y$ are unrelated too. Therefore, $\range{U|Y(\omega)=y}=\range{U}$. This shows that $\mathcal{L}(U\rightarrow Y)=0$.
\end{IEEEproof}

For the Markov chain $U-X-Y$, the measure $\mathcal{L}(U \rightarrow Y)$ can be used to quantify the non-stochastic brute-force guessing leakage if we know attribute $U$ of $X$ that is targeted by the adversary. However, there are some real-world situations that we do not know \emph{a priori} the intention of the adversary, i.e., the attribute $U$ of $X$ that the adversary is trying to infer. In some cases, more than one user may observe $Y$ and each user might be interested in guessing/estimating a different attribute of $X$. In these situations, it is required to consider the brute-force guessing leakage $\mathcal{L}(U \rightarrow Y)$ when the attribute $U$ varies. Therefore, we need to define a {maximal non-stochastic} guessing leakage. This is in-line with axiom R2 in~\cite{issa2018operational}. We consider such situations in the next section.

\section{Maximal Non-Stochastic Leakage} \label{sec:maximal_leakage}
For given uv $X$ and the released output $Y$, we define the maximal non-stochastic brute-force guessing leakage over all attributes $U$ as follows.




\begin{definition}[\underline{Maximal Non-Stochastic Brute-Force Leakage}]
The maximal non-stochastic leakage from $X$ to $Y$ is defined as
\begin{align}\label{eqn:maximal_leakage}
\mathcal{L}_\star(X\rightarrow Y)=&\sup_{U \colon U-X-Y}\mathcal{L}(U\rightarrow Y),
\end{align}
where the supremum is taken over all {functions} $g:\range{X}\rightarrow\range{U}$ with $\range{U}$ containing finite arbitrary alphabets.
\end{definition}




The maximal non-stochastic brute-force leakage only depends on uvs $X$ and $Y$. The maximizer of \eqref{eqn:maximal_leakage} denotes the most vulnerable attributes $U$ to the brute-force guessing over $\range{Y|X}$\Ni{;} The supremum of \eqref{eqn:maximal_leakage} indicates the lowest data privacy level the channel $\range{Y|X}$ provides.

Now, we can show that maximal non-stochastic leakage admits axiom R3 in the axiomatic approach to operational information leakage in~\cite{issa2018operational}. That is, maximal non-stochastic leakage satisfies data processing inequality (post processing does not increase leakage), independence property (statistically independent outputs result in zero leakage), and additivity property.

\begin{proposition}[\underline{Properties of Maximal Leakage}] The following holds:
\begin{itemize}
\item[a)] $\mathcal{L}_\star(X\rightarrow Y)\geq 0$;
\item[b)] $\mathcal{L}_\star(X\rightarrow Y)=0$ if and only if $X$ is unrelated to $Y$;
\item[c)] $\mathcal{L}_\star(X\rightarrow Y)\leq H_0(X)$ with the equality if $Y = X$;
\item[d)] 
$\mathcal{L}_\star(X\rightarrow Z)\leq \mathcal{L}_\star(X\rightarrow Y)$ if Markov chain $X-Y-Z$ holds;
\item[e)] If $(X_i,Y_i)$, $\forall i$, are unrelated, i.e., $(X_i,Y_i)$ and $(X_{i'},Y_{i'})$ are unrelated $\forall i\neq i'$, then 
\begin{align*}
    \mathcal{L}_\star((X_1,\dots,X_n)\rightarrow (Y_1,\dots,Y_n))
    =\sum_{i=1}^n\mathcal{L}(X_i\rightarrow Y_i).
\end{align*}
\end{itemize}
\end{proposition}

\begin{IEEEproof} Proof of (a): Note that $\mathcal{L}(U \rightarrow Y)\geq 0$ for all $U$ such that $U-X-Y$ is a Markov chain; see Proposition~\ref{prop:Bounding_Leakage}.
Taking maximum of both sides of this inequality results in (a).

%
Proof of (b): According to Proposition~\ref{prop:Bounding_Leakage}, for unrelated $X$ and $Y$, $\mathcal{L}(U \rightarrow Y) = 0$ for all $U$ such that $U-X-Y$ is a Markov chain. Hence, $\mathcal{L}_\star(X \rightarrow Y) = 0$. Now, we prove the reverse. Assume that $\mathcal{L}_\star(X \rightarrow Y) = 0$. This implies that $\mathcal{L}(U \rightarrow Y) = 0$ for all $U$ such that $U-X-Y$ is a Markov chain. For the special case that $U = X$,
 $\mathcal{L}(U \rightarrow Y)=\mathcal{L}(X \rightarrow Y)={\max_{y\in\range{Y}} \log_2 (|\range{X}|/ |\range{X|Y(\omega)=y}|)} = 0$ and hence we must have $|\range{X|Y(\omega)=y}| = |\range{X}|$ 
for all $y\in\range{Y}$. Noting that $\range{X|Y(\omega)=y}\subseteq\range{X}$ {and therefore} $|\range{X|Y(\omega)=y}| = |\range{X}|$ implies that $\range{X|Y(\omega)=y}= \range{X}$. Hence, $X$ and $Y$ must be unrelated.

Proof of (c): Notice that we have $\mathcal{L}(U\rightarrow Y)=\max_{y\in \range{Y}} \log_2
\left(|\range{U}|/|\range{U|Y(\omega)=y}|\right)\leq \log_2(|\range{U}|)$ because $|\range{U|Y(\omega)=y}|\geq 1$. Further, we have $|\range{U}|\leq |\range{X}|$. Hence, $\mathcal{L}(U\rightarrow Y)\leq \log_2(|\range{X}|)=H_0(X)$ for all $U$. Taking maximum of left hand side of this inequality over all $U$ results in (c). For $Y=X$, $\mathcal{L}(U\rightarrow Y)={ \mathcal{L}(U\rightarrow X) =}\max_{x\in \range{X}} \log_2
\left(|\range{U}|/|\range{U|X(\omega)=x}|\right)$. Note that $\range{U|X(\omega)=x}=\{g(x)\}$ is a singleton and, as a result, $|\range{U|X(\omega)=x}|=1$. This implies that $\mathcal{L}(U\rightarrow Y)=|\range{U}|$. Further, $|\range{U}|\leq |\range{X}|$ with equality achieved if $U=X$. Thus,
$\mathcal{L}_\star(X\rightarrow Y)=\sup_{U\colon U-X-Y}\mathcal{L}(U\rightarrow Y)=H_0(X).$

Proof of (d): For $U$ that holds Markov Chain $U-X-Y-Z$, we have $\mathcal{L}(U\rightarrow Y) \geq \mathcal{L}(U\rightarrow Z)$. Taking maximum of both sides of this inequality results in (d).


Proof of (e): We have $\mathcal{L}((U_i)_{i=1}^n\rightarrow (Y_i)_{i=1}^n)=\sum_{i=1}^n \mathcal{L}(U_i\rightarrow Y_i)$ if $(U_i,X_i,Y_i)$, $\forall i$, are unrelated~\cite{farokhi2019noiseless}. Note that, by definition, $(U_i,X_i,Y_i)$, $\forall i$, are unrelated if $(X_i,Y_i)$, $\forall i$, are unrelated. Taking maximum from both sides of this equality over $(U_i)_{i=1}^n$, such that $(U_i)_{i=1}^n-(X_i)_{i=1}^n-(Y_i)_{i=1}^n$ forms a Markov chain, proves (e).
\end{IEEEproof}


Now, we are ready to present a formula for computing the maximal non-stochastic leakage. This is done in the next proposition.

\begin{proposition}[\underline{Computing Maximal Leakage}] \label{prop:Bounds_on_Maximal_Leakage} $\mathcal{L}_\star(X\rightarrow Y)=\log_2(|\range{X}| - \min_{y\in\range{Y}} |\range{X|Y(\omega)=y}| + 1)$.
\end{proposition}

\begin{IEEEproof}
We start by proving that $\mathcal{L}_\star(X\rightarrow Y)\leq \log_2(|\range{X}| - \min_{y\in\range{Y}} |\range{X|Y(\omega)=y}| + 1)$. To do so, we need to prove that, $\forall y\in\range{Y}$,
\begin{align}\label{eqn:middle_result}
|\range{X}| - |\range{U}| \geq |\range{X|Y(\omega)=y}| - |\range{U|Y(\omega)=y}|.
\end{align}
This is done by \textit{reductio ad absurdum}. Assume that~\eqref{eqn:middle_result} does not hold for all $y\in\range{Y}$. Therefore, there must exists $y\in\range{Y}$ such that
\begin{align} \label{eqn:reductio_ad_absurdum}
|\range{X}| - |\range{U}| < |\range{X|Y(\omega)=y}| - |\range{U|Y(\omega)=y}|,
\end{align}
Subtracting $|\range{U}\setminus \range{U|Y(\omega)=y}|$ from both sides of~\eqref{eqn:reductio_ad_absurdum} results in
\begin{align*}
|\range{X}| - |\range{U}|-|\range{U}&\setminus \range{U|Y(\omega)=y}| \\
<& |\range{X|Y(\omega)=y}| - |\range{U|Y(\omega)=y}|\\
&-|\range{U}\setminus \range{U|Y(\omega)=y}|\\
=& |\range{X|Y(\omega)=y}| - |\range{U}|,
\end{align*}
where the equality follows from that $|\range{U}|=|\range{U|Y(\omega)=y}|+|\range{U}\setminus \range{U|Y(\omega)=y}|$ because $(\range{U}\setminus \range{U|Y(\omega)=y})\cap \range{U|Y(\omega)=y}=\emptyset$ and $(\range{U}\setminus \range{U|Y(\omega)=y})\cup \range{U|Y(\omega)=y}=\range{U}$. Therefore, it must be that
\begin{align*}
|\range{X}|-|\range{U}\setminus \range{U|Y(\omega)=y}|< |\range{X|Y(\omega)=y}|.
\end{align*}
or equivalently
\begin{align*}
|\range{X}|- |\range{X|Y(\omega)=y}|<|\range{U}\setminus \range{U|Y(\omega)=y}|.
\end{align*}
Because $(\range{X}\setminus\range{X|Y(\omega)=y})\cap \range{X|Y(\omega)=y}=\emptyset$ and ${\range{X}=}(\range{X}\setminus\range{X|Y(\omega)=y})\cup \range{X|Y(\omega)=y}$, we have $|\range{X}|=|\range{X}\setminus\range{X|Y(\omega)=y}|+|\range{X|Y(\omega)=y}|$.
Therefore, it must be that
\begin{align} \label{eqn:middle_result:1}
|\range{X}\setminus\range{X|Y(\omega)=y}|<|\range{U}\setminus \range{U|Y(\omega)=y}|.
\end{align}
On the other hand, we have
\begin{equation}
     \begin{aligned} \nonumber
         &|\range{U}\setminus \range{U|Y(\omega)=y}| \\
         & \qquad\quad = \big| \{g(x) \colon x \in \range{X} \} \setminus \{g(x) \colon x \in \range{X|Y(\omega)=y}\} \big| \\
         & \qquad\quad  \leq \big| \{g(x) \colon x \in \range{X} \setminus \range{X|Y(\omega)=y} \} \big| \\
         & \qquad\quad \leq \big| \range{X} \setminus \range{X|Y(\omega)=y} \big|,
     \end{aligned}
\end{equation}
which contradicts \eqref{eqn:middle_result:1}. Thus,~\eqref{eqn:middle_result} must be valid for all $y\in\range{Y}$.
%

Using~\eqref{eqn:middle_result}, we get
    \begin{equation} \nonumber
        \begin{aligned}
            \frac{|\range{U}|}{|\range{U|Y(\omega)=y}|}
            & \leq \frac{|\range{X}| - |\range{X|Y(\omega)=y}|}{|\range{U|Y(\omega)=y}|} + 1 \\
            & \leq |\range{X}| - |\range{X|Y(\omega)=y}| + 1, \; \forall y\in\range{Y},
        \end{aligned}
    \end{equation}
    where the last inequality holds because
    \begin{align}
    |\range{U|Y(\omega)=y}|
    &=\left|\bigcup_{x\in\range{X|Y(\omega)=y}}\range{U|X(\omega)=x}\right|\geq 1.\label{eqn:proof:3}
    \end{align}
    Using $y^* \in \argmin_{y\in\range{Y}} |\range{U |Y(\omega)=y}|$, we get
        \begin{align}
            \mathcal{L}(U\rightarrow Y) &= \log_2\left( \frac{|\range{U}|}{|\range{U|Y(\omega)=y^*}|}\right)  \nonumber\\
            & \leq \log_2 (|\range{X}| - |\range{X|Y(\omega)=y^*}| + 1) \nonumber\\
            & =\log_2 (|\range{X}|\hspace{-.03in} -\hspace{-.03in} \min_{y\in\range{Y}}|\range{X|Y(\omega)=y}| + 1).\label{eq:LUBaux}
        \end{align}
    Since inequality \eqref{eq:LUBaux} holds for all $U$, we have the proved upper bound.

    Now, we continue by proving the lower bound that $\mathcal{L}_\star(X\rightarrow Y)\geq \log_2(|\range{X}|-\min_{y\in\range{Y}}|\range{X|Y(\omega)=y}|+1)$. Select an arbitrary $y^*\in \argmin_{y\in\range{Y}}|\range{X|Y(\omega)=y}|$. Let us define two sets $\mathcal{X}_1:=\range{X|Y(\omega)=y^*}$ and $\mathcal{X}_2:=\range{X}\setminus\mathcal{X}_1$. Define
    $g:\range{X}\rightarrow\range{U}$ with $\range{U}=\mathcal{X}_2\cup\{u^*\}$ as
    \begin{align} \label{eq:WorstCaseAttri}
        g(x)
        =
        \begin{cases}
            u^*, & x\in\mathcal{X}_1,\\
            x, & x\in\mathcal{X}_2.
        \end{cases}
    \end{align}
    Note that, by construction, $|\range{U|Y(\omega)=y^*}|=|\{u^*\}|=1$ and $|\range{U|Y(\omega)=y}|=|g(\range{X|Y(\omega)=y})|\geq 1$ for all $y\in\range{Y}\setminus\{y^*\}$. Hence, $\min_{y\in\range{Y}} |\range{U|Y(\omega)=y}|=1$. Therefore,
    \begin{align*}
        \mathcal{L}_\star(X\rightarrow Y)
        &\geq \mathcal{L}(U\rightarrow Y)\\
        &=\log_2\left( \frac{|\range{U}|}{ \displaystyle\min_{y \in \range{Y}} |\range{U|Y(\omega)=y}|}\right) \\
        &=\log_2(|\range{U}|)\\
        &=\log_2(|\range{X}\setminus \range{X|Y(\omega)=y^*}|+1)\\
        &=\log_2(|\range{X}|-|\range{X|Y(\omega)=y^*}|+1)\\
        &=\log_2(|\range{X}|-\min_{y\in\range{Y}}|\range{X|Y(\omega)=y}|+1).
    \end{align*}
    This concludes the proof.
\end{IEEEproof}

\begin{remark}
    The function $g$ in~\eqref{eq:WorstCaseAttri} constructs the most vulnerable attribute $U$ of uv $X$, which is determined by any ${y^* \in }\argmin_{y\in\range{Y}}|\range{X|Y(\omega)=y}|$.
\end{remark}


\begin{corollary} $\mathcal{L}_\star(X\rightarrow Y)$ is not symmetric in general.
\end{corollary}

\begin{IEEEproof} For uvs $X$ and $Y$ with joint range $\range{X,Y}=\{(x_1,y_1),(x_2,y_1),(x_3,y_2)\}$, we have  $\mathcal{L}_\star(X\rightarrow Y)=\log_2(3)\neq \log_2(2)=\mathcal{L}_\star(Y\rightarrow X)$.
\end{IEEEproof}

In the next section, we introduce non-stochastic identifiability as a new notion of privacy, motivated by the expression for the maximal leakage in Proposition~\ref{prop:Bounds_on_Maximal_Leakage}.



\section{Non-Stochastic Identifiability} \label{sec:privacy}
We define non-stochastic identifiability by requiring that the ratio of the cardinality of the set of compatible realization of uv $X$ with access to the measurements of uv $Y$ over the cardinality of the set of compatible realization of uv $X$ without this auxiliary information is lower bounded by an exponential of the privacy budget. This implies that access to the realizations of $Y$ does not significantly reduce the cardinality of the set of possibilities that must be tested for guessing the realization of $X$. 
%
This definition is in consistent with stochastic identifiability in~\cite{wang2016relation,lee2012differential} which requires that the posterior distribution (instead of the conditional range) to remain similar with and without access to privacy-preserving measurements.

\begin{definition}[Non-Stochastic Identifiability] Any mapping $\mathfrak{M}$ is $\epsilon$-identifiable, for $\epsilon>0$, if
\begin{align}
    |\range{X|Y(\omega)=y}|\geq |\range{X}|2^{-\epsilon}, {\quad \forall y \in \range{Y}},
\end{align}
with $Y=\mathfrak{M}\circ X$.
\end{definition}

We refer to $\epsilon$ in the non-stochastic identifiability as {the} \emph{privacy budget}. By decreasing the privacy budget, we ensure a higher level of privacy (cf., differential privacy~\cite{dwork2006calibrating} and identifiability~\cite{wang2016relation}). This is intuitively because, by decreasing the privacy budget, the size of the set $\range{X|Y(\omega)=y}$ increases and thus guessing the actual realization of uv $X$ becomes more complex.

\begin{corollary} \label{cor:ident} For any $\epsilon$-identifiable mapping $\mathfrak{M}$, $\mathcal{L}_\star(X\rightarrow Y)\leq \log_2(|\range{X}|(1-2^{-\epsilon}) + 1)$.
\end{corollary}

\begin{IEEEproof}
The proof follows from that
$\mathcal{L}_\star(X\rightarrow Y)
    =\log_2(|\range{X}| - \min_{y\in\range{Y}} |\range{X|Y(\omega)=y}| + 1)
    \leq \log_2(|\range{X}|(1-2^{-\epsilon}) + 1)$.
\end{IEEEproof}

Corollary~\ref{cor:ident} shows that, as expected, the maximal non-stochastic brute-force guessing leakage $\mathcal{L}_\star(X\rightarrow Y)$ goes to zero as the privacy budget approaches zero. By increasing the privacy budget, however, we increase the bound on the maximal non-stochastic brute-force guessing leakage $\mathcal{L}_\star(X\rightarrow Y)$ and therefore more private information could be potentially leaked.

\section{Brute-Force to One-Shot Guess} \label{sec:relationship}
In the previous section, we considered a brute-force guessing adversary that can potentially check all the possibilities of the private information in $\range{U|Y(\omega)=y}$ that are compatible with the available outputs $Y(\omega)=y$ of the channel $\range{Y|X}$ to find the actual private realization. In this section, we restrict ourselves to one-shot guesses. We first analyze the non-stochastic case and its relationship with the non-stochastic brute-force guessing.

\subsection{Non-Stochastic One-Shot Guessing}
Let us consider an adversary with only a single opportunity for guessing the private realization of uv $U$ by observing the realization of uv $Y$. For instance, consider the problem of guessing a person's password based on side-channel information (e.g., inter-keystroke delay as in~\cite{issa2018operational}) while the system locks immediately after one wrong guess. Therefore, the adversary is interested in finding the largest amount of information that can be deduced correctly with one guess. This happens when $|\range{U|Y(\omega)=1}|=1$ for all $y\in\range{Y}$. In the next proposition, we show that the maximum information is the largest amount of information can be leaked to such an adversary. We further relate this notion of leakage to maximal non-stochastic leakage with brute-force guessing.

\begin{proposition}[\underline{Maximal Leakage Bounds Maximin Info}] \label{prop:maximin}
For uvs $X$ and $Y$,
\begin{align*}
    I_\star(X;Y)=
    \hspace{-.1in}
    \sup_{
    \scriptsize
    \begin{array}{c}
    \scriptsize U\colon U-X-Y, \\
    \scriptsize |\range{U|Y(\omega)=y}|=1, \\
    \forall y\in\range{Y}
    \end{array}
    }
    \hspace{-.1in}\mathcal{L}(U\rightarrow Y)
    \leq
    \mathcal{L}_\star(X\rightarrow Y),
\end{align*}
where the supremum is taken over  all $g:\range{X}\rightarrow\range{U}$ such that $|\{g(x) \colon x \in \range{X | Y(\omega)=y}\}| =  |\range{U|Y(\omega)=y}|=1$.
\end{proposition}

\begin{IEEEproof}
	The second inequality trivially follows from that increasing the search domain of the supermum operator results in a larger value. Therefore, we only focus on the first inequality. Note that $|\range{U|Y(\omega)=y}|=1$ implies that there exists $f$ such that $U=f(Y)$. Therefore, $U=f(Y)=g(X)$. Following Lemma 1 in~\cite{wolf2004zero}, we know that there exists a function $h$ such that $U=h(X\wedge Y)$, where $X\wedge Y$ is the common variable in the sense of~\cite{wolf2004zero} defined for uncertain variables (instead of random variables) following the approach of~\cite{mahajanrelationship}. Therefore,
	$\mathcal{L}(U\rightarrow Y)=H_0(U)\leq H_0(X\wedge Y)=I_\star(X;Y).$
	Since this inequality holds for all $U$ such that $\range{U|Y(\omega)=y}=1$, we get
	\begin{align*}
	    \sup_{
    \scriptsize
    \begin{array}{c}
    \scriptsize U\colon U-X-Y, \\
    \scriptsize |\range{U|Y(\omega)=y}|=1, \\
    \forall y\in\range{Y}
    \end{array}
    }\mathcal{L}(U\rightarrow Y)\leq
	    I_\star(X;Y).
	\end{align*}
 	{On the other hand, for $U^* = X \wedge Y$,
	\begin{align*}
	    \sup_{
    \scriptsize
    \begin{array}{c}
    \scriptsize U\colon U-X-Y, \\
    \scriptsize |\range{U|Y(\omega)=y}|=1, \\
    \forall y\in\range{Y}
    \end{array}
    }\hspace{-.2in}\mathcal{L}(U\rightarrow Y)
    \geq \mathcal{L}(U^*\rightarrow Y)=I_\star(X;Y).
	\end{align*}
	}
	Combining these inequalities concludes the proof.
\end{IEEEproof}


\begin{remark}[Relationship with Zero-Error Capacity]
Following Proposition~\ref{prop:maximin} and~\cite{nair2013nonstochastic}, the zero-error capacity of any memoryless uncertain channel satisfies $C_0=\sup_{\range{X}\subseteq\mathbb{X}}I_\star (X;Y)\leq  \sup_{\range{X}\subseteq\mathbb{X}} \mathcal{L}_\star(X\rightarrow Y).$
Therefore, based on Corollary~\ref{cor:ident}, the zero-error capacity of any memoryless $\epsilon$-identifiable channel is upper bounded by $\log_2(|\mathbb{X}|(1-2^{-\epsilon}) + 1)$, where $|\mathbb{X}|$ is the number of the input alphabets.
This constraints dynamical systems that can be estimated or stabilized through privacy-preserving communication channels~\cite{nair2013nonstochastic, matveev2007shannon}.
\end{remark}


In the next subsection, we consider one-shot guessing in the stochastic sense of~\cite{issa2018operational} and investigate its relationship with the maximal non-stochastic leakage with brute-force guessing.

\subsection{Maximal Stochastic Leakage}
We can recreate the stochastic framework for information leakage in~\cite{issa2018operational} by endowing all the uncertain variables in this paper with a measure. 



\begin{definition}[\underline{Maximal Stochastic Leakage}]
	For jointly distributed rvs $X$ and $Y$, the maximal stochastic leakage from $X$ to $Y$ is given by
	\begin{align*}
	\widetilde{\mathcal{L}}(X&\rightarrow Y)\\
	&=
	\sup_{U\colon U-X-Y} \log_2\left( \frac{\displaystyle\mathbb{E}\left\{\max_{u \in \range{U}} \mathbb{P}\{U=u|Y=y\}\right\}}{\displaystyle\max_{u\in\range{U}}\mathbb{P}\{U=u\}}\right),
	\end{align*}
	where supremum is taken over all {random variables (rvs)} $U$ taking values in finite arbitrary alphabets. It was shown in~\cite{issa2018operational} that
	\begin{align*}
	\widetilde{\mathcal{L}}(X\rightarrow Y)
	&=\log_2\left(\sum_{y\in\range{Y}}\max_{x\in\range{X}}\mathbb{P}\{Y=y|X=x\}\right)\\
	&=I_{\infty} (X;Y),
	\end{align*}
	where $I_{\infty}$ is the Sibson mutual information $I_{\alpha}$ in the order $\alpha \rightarrow \infty$~\cite{sibson_information_1969, verdu_alpha_mutual_2015}. Note the fact that $\{x\colon \mathbb{P}\{X=x\}>0\} = \range{X}$.
\end{definition}

In the next proposition, we show that the worst-case maximal stochastic leakage provides a bound for the maximal non-stochastic brute-force leakage. {Therefore, we can interpret the maximal non-stochastic brute-force leakage as a robust  non-stochastic counterpart of the maximal stochastic leakage.} 

\begin{proposition}[\underline{Relating Maximal Leakages}] $\mathcal{L}_\star(X\rightarrow Y)\leq \sup_{\mathbb{P}\{Y=y|X=x\}}\widetilde{\mathcal{L}}(X\rightarrow Y)+H_0(X|Y)$.
\end{proposition}
\begin{IEEEproof} We start by proving that $\mathcal{L}_\star(X\rightarrow Y)\leq H_0(Y)+H_0(X|Y)$. First, note that
\begin{align*}
\range{U}
=&\bigcup_{y'\in\range{Y}}\bigcup_{x'\in\range{X|Y(\omega)=y'}}\range{U|Y(\omega)=y',X(\omega)=x'}\\
=&\bigcup_{y'\in\range{Y}}\bigcup_{x'\in\range{X|Y(\omega)=y'}}\range{U|X(\omega)=x'},
\end{align*}
because $U-X-Y$ forms a Markov uncertainty chain. As a result,
\begin{align}
|\range{U}|& =\Bigg|\bigcup_{y'\in\range{Y}}\bigcup_{x'\in\range{X|Y(\omega)=y'}}\range{U|X(\omega)=x'}\Bigg|\nonumber\\
&\leq |\range{Y}|\max_{y\in\range{Y}}|\range{X|Y(\omega)=y}| ,\label{eqn:proof:4}
\end{align}
because $|\range{U|X(\omega)=x}|=1$ for all $x\in\range{X}$. Combining~\eqref{eqn:proof:3} and~\eqref{eqn:proof:4} results in
\begin{align*}
\frac{|\range{U}|}{\displaystyle\min_{y\in \range{Y}} |\range{U|Y(\omega)=y}|}
&\leq |\range{Y}|\max_{y\in\range{Y}}|\range{X|Y(\omega)=y}|.
\end{align*}
Therefore, $\mathcal{L}(X\rightarrow Y)\leq H_0(Y)+H_0(X|Y)$ for all $U$. This proves the upper bound $\mathcal{L}_\star(X\rightarrow Y)\leq H_0(Y)+H_0(X|Y)$.
The rest of the proof follows from that $\sup_{\mathbb{P}\{Y=y|X=x\}}\widetilde{\mathcal{L}}(X\rightarrow Y)=H_0(Y)$ because of~\cite[Lemma~1 \& Example~6]{issa2018operational}.
\end{IEEEproof}

\section{Conclusions and Future Work} \label{sec:conc}
We developed an interpretable notion of non-stochastic information leakage based on guessing in a non-stochastic framework. We considered brute-force guessing in which an adversary can potentially check all the possibilities of the private information that are compatible with the available outputs to find the actual private realization. The ratio of the worst-case number of guesses for the adversary in the presence of the output and in the absence of it captures the reduction in the adversary's guessing  complexity and is thus used as a measure of information leakage. We computed the maximal non-stochastic leakage {over all sensitive attributes that could be targeted by the adversary and compared it with non-stochastic identifiabiliy, maximin information, and stochastic maximal leakage}. Future work can focus on extending this definition to a dynamic framework with continual observations.

\bibliography{citation}
\bibliographystyle{ieeetr}

\end{document}